\documentclass[final,1p,times]{elsarticle}

\usepackage{amssymb}
\usepackage{xfrac}
\usepackage{amsmath}

\usepackage{lineno}

\journal{Journal of Sound and Vibration}

\begin{document}

\begin{frontmatter}



\title{Energy harvesting below the onset of flutter}


\author[dyn]{Merten Stender\corref{cor1}}
\ead[dyn]{m.stender@tuhh.de}
\author[audi]{Merten Tiedemann}
\author[dyn,ic]{Norbert Hoffmann}

\address[dyn]{Hamburg University of Technology, Hamburg, Germany}
\address[audi]{AUDI AG, Ingolstadt, Germany}
\address[ic]{Imperial College London, London, United Kingdom}

\cortext[cor1]{Corresponding author}

\begin{abstract}

This work demonstrates preliminary results on energy harvesting from a linearly stable flutter-type system with circulatory friction forces. Harmonic external forcing is applied to study the energy flow in the steady sliding configuration. In certain parameter ranges negative excitation work is observed where the external forcing allows to pull part of the friction energy out of the system and thus makes energy harvesting possible. Studies reveal that this behavior is largely independent of the flutter point and thus that it is primarily controlled by the excitation. Contrary to existing energy harvesting approaches for such systems, this approach uses external forcing in the linearly stable regime of the oscillator which allows to control vibrations and harvest energy on demand. 
\end{abstract}


\end{frontmatter}


\section{Introduction}

Energy harvesting from vibrations \cite{Kazmierski.2011, Wei.2017} has recently become an important research field driven by the development of micro-electromechanical systems. When harvesting energy from ambient vibrations, the central limitation of the energy conversion is the frequency range of the resonator compared to the spectrum of the vibrations to be harvested. The energy efficiency decreases rapidly when the frequency spectrum of the vibrations shifts out of the resonator's optimal point of operation. Energy harvesting from high-amplitude friction-induced vibrations (FIV) was proposed recently \cite{Wang.2018b}. Even though friction-excited systems are well known to exhibit rich periodic and chaotic dynamic behavior \cite{Stender.2019c}, most systems show high vibration energy shares in narrow frequency ranges. Hence, energy harvesting resonators can be tuned accordingly and help to overcome the limitations of suboptimal frequency range coverage. However, those harvesting approaches rely on the existence of high-amplitude vibrations - which in most cases is an unwanted dynamic state. Loosely speaking, energy harvesting from FIV '\textit{makes the best out of an undesirable situation}'. Furthermore, considering real-world FIV, the energy generation remains mostly uncontrollable and unpredictable due to their inherently fugitive and irregular nature. 

External forcing of self-excited friction systems has been mostly studied as active damping measure \cite{Hoffmann.2005}. Additionally, forcing has been applied in various ways to predict the stability behavior of friction systems with flutter-type, i.e. mode-coupling, instabilities. Schlagner and von Wagner~\cite{Schlagner.2007} examine various friction oscillator models under external forcing of the normal contact force. The authors find that the phase shift between the in-plane vibration and the friction force controls the friction work put into or pulled out of the system, and thus stability. Experimentally, the modal properties and friction forces of a full brake system are analyzed in \cite{Stegmann.2015} to assess the system robustness. Higher friction energy inputs indicated configurations that were prone to show FIV. A recent study \cite{Tiedemann.2015c} investigates the behavior of a minimal nonlinear self-excited system under harmonic forcing. The amplitudes of the frequency response function (FRF) and the work done by circulatory forces are investigated as potential predictors for instability. Locally, high values of those metrics are observed before the point of instability. 
 
In the following, we illustrate how external forcing of a linear friction oscillator can be used to harvest energy. For this purpose we study the work done by damping, circulatory and external forces in the linearly stable regime of the system. 

\section{Linear flutter-type system}

\begin{figure}[h!]
\includegraphics[width=1.0\columnwidth]{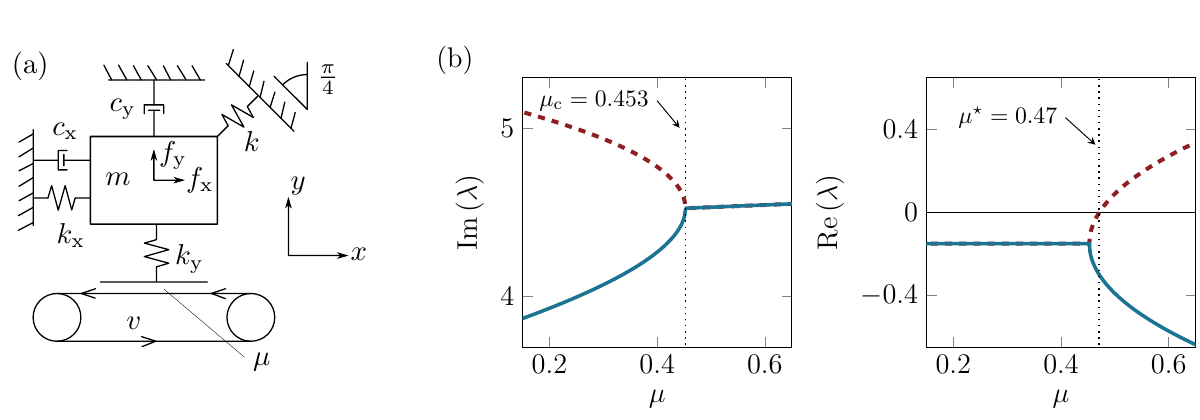}%
\caption{(a) linear 
friction oscillator \cite{Hoffmann.2003} and (b) portrait of the eigenvalues $\lambda$ exhibiting flutter-type instability for a variation of the friction coefficient $\mu$ with $m=1$\,kg, $k_{\mathrm{x}}=11$\,$\mathrm{\sfrac{N}{m}}$, $k_{\mathrm{y}}=20$\,$\mathrm{\sfrac{N}{m}}$, $k=5$\,$\mathrm{\sfrac{N}{m}}$, $c_{\mathrm{x}}=0.3$\,$\mathrm{\sfrac{Ns}{m}}$, $c_{\mathrm{y}}=0.3$\,$\mathrm{\sfrac{Ns}{m}}$, $v>\dot{x}\left(t\right)$ }%
\label{fig_01}%
\end{figure}

We are considering the dynamic system \cite{Hoffmann.2003} in Figure~\ref{fig_01}~(a) 
\begin{equation}
\mathbf{M} \ddot{\mathbf{x}} + \mathbf{D} \dot{\mathbf{x}} + \left(\mathbf{K}+\mathbf{N} \right) \mathbf{x} = \mathbf{f}_{\mathrm{exc}}\left( t \right) \quad \mathbf{M}, \mathbf{D}, \mathbf{K}, \mathbf{N} \in \mathbb{R}^{2,2} \quad \text{(see \ref{EoM}) }
\label{eq:EoM_general}
\end{equation}
with energy input from the friction interface and external forcing $\mathbf{f}_{\mathrm{exc}}$. Motions in horizontal and vertical directions are viscously damped and coupled through the inclined stiffness $k$. The speed of the conveyor belt is assumed to always be larger than the horizontal velocity of the slider such that the system is not undergoing any stick-slip type dynamics and can be taken as linear. This assumption for high relative sliding velocities is supported by a recent experimental study \cite{Stender.2019c}. A Coulomb-type friction force creates the asymmetrical circulatory terms $\mathbf{N}=-\mathbf{N}^\top$ that allow the system to flutter \cite{Hoffmann.2003}. This mode-coupling behavior is depicted in Figure~\ref{fig_01}~(b). At the coupling point $\mu_{\mathrm{c}}=0.453$, the first mode is uni-directionally forcing the second mode at resonance frequency. Once this mode experiences negative damping at $\mu^{\star}=0.47$, the system becomes linearly unstable. For damped systems the points of coupling and instability are usually not identical. 

\section{Negative excitation work}

We study the system in the strictly stable configuration, i.e. $\mu < \mu^{\star} = 0.47$. The balance of work $W$ done in the forced system

\begin{equation}
\underbrace{ \int_{0}^{\frac{2 \pi}{\Omega}}{\dot{\mathbf{x}}^\top \mathbf{f}_{\mathrm{exc}}\,\mathrm{d} t}}_{W_{\mathrm{exc}}} \hspace{1mm}-\hspace{1mm}\underbrace{\int_{0}^{\frac{2 \pi}{\Omega}}{\dot{\mathbf{x}}^\top \mathbf{C} \dot{\mathbf{x}}\,\mathrm{d} t}}_{W_{\mathrm{damp}}} \hspace{1mm} - \hspace{1mm}\underbrace{ \int_{0}^{\frac{2 \pi}{\Omega}}{\dot{\mathbf{x}}^\top \mathbf{N} \dot{\mathbf{x}}\,\mathrm{d} t}}_{W_{\mathrm{circ}}} \hspace{1mm}= 0
\label{eq:EnergyBalance}
\end{equation}

is obtained by integrating the respective power over one period of vibration. The terms of inertia and stiffness vanish due to their conservative character and the viscous dampers represent energy sinks. In the following, we consider a horizontal forcing.

\begin{figure}[h!]
\includegraphics[width=1.0\columnwidth]{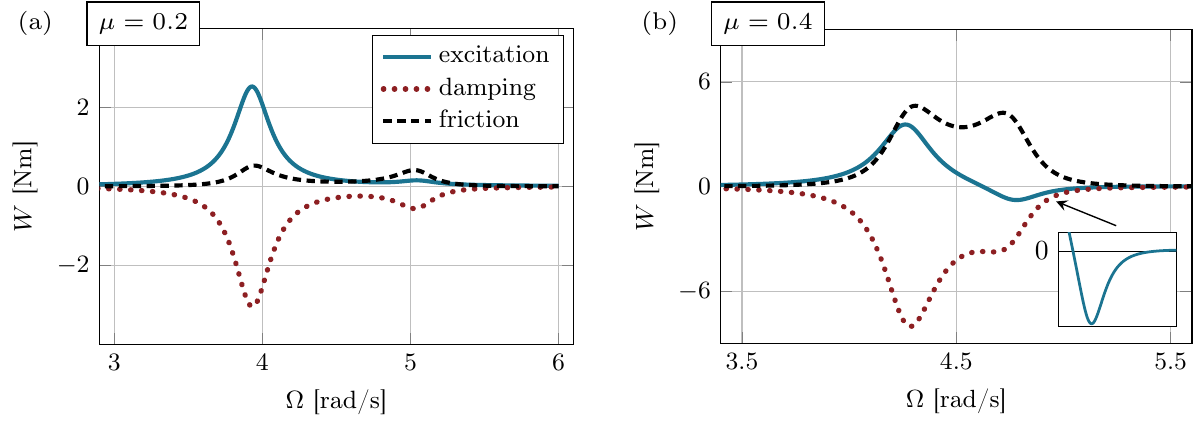}%
\caption{Work done in the forced system for $\mathbf{f}_{\mathrm{exc}}=\left[1, 0 \right]^\top \cdot \cos \left( \Omega t \right)$ in the stable regime. Negative excitation work is found in (b) around the second mode $4.62 \leq \Omega \leq 5.34$\,rad/s for higher friction values}%
\label{fig_02}%
\end{figure}

The work at $\mu=0.2$ is displayed in Figure~\ref{fig_02}~(a) for an excitation frequency sweep. The first and second mode are clearly visible at $\Omega=3.9$\,rad/s and $\Omega=5.0$\,rad/s. The energy put into the system by the friction and the excitation is dissipated in the dampers to maintain the energy balance of the system. When the friction value is increased to $\mu=0.4$, an interesting observation can be made, see Figure~\ref{fig_02}~(b): While the balance of work does not qualitatively change for the first mode, the excitation work changes sign close to the second mode. For this configuration the system exhibits negative excitation work when forced in the frequency range of $4.62 \leq \Omega \leq 5.34$\,rad/s. Here, the external forcing in fact pulls energy out of the system. As a result, the overall energy flow changes from one sink, i.e. the dampers, to two sinks, i.e. the dampers and the excitation. At the second mode $82\%$ of the total friction energy input are dissipated in the dampers and $18\%$ are pulled out of the system by the forcing, resulting in the net efficiency of $\xi = \sfrac{W_{\mathrm{in}}}{W_{\mathrm{out}}}=0.18$. This means that the forcing channels energy stemming from the friction interface out of the system, and thus allows for energy harvesting. The system seems to develop its own dynamics counteracting the forcing. Most importantly, the harvesting is operated in the stable regime of the linear oscillator. Thus, if the forcing is turned off, the system will return to its equilibrium sliding state. This behavior is fundamentally different to existing approaches, where high-amplitude vibrations, i.e. in the linearly unstable regime of the system, are utilized for energy harvesting. To actually design such an energy harvester, one will have to take into consideration the required power electronics, resulting electro-mechanical interactions and the resulting decrease in the effective conversion rate $\xi$. \\

\begin{figure}[h!]
\centering
\includegraphics[width=1.0\columnwidth]{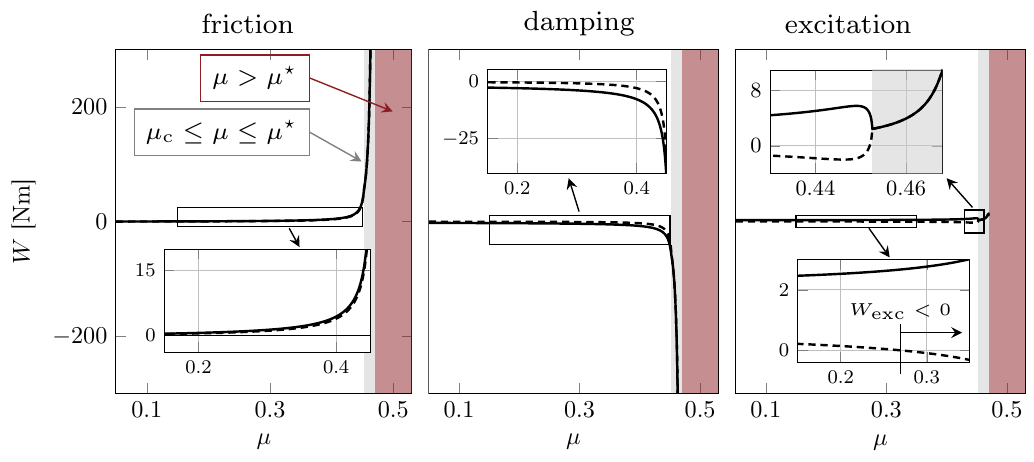}%
\caption{Work along the friction coefficient for excitation frequencies at the first (solid line) and second mode (dashed line), i.e. $\Omega_{1,2}=\mathrm{Im} \left(\lambda_{1,2} \right)$, $\mathbf{f}_{\mathrm{exc}}=\left[1, 0\right]^\top \cdot \cos \left( \Omega_{1,2} t \right)$. Negative excitation work is observed at the second mode for $0.261 \leq \mu \leq 0.452$, i.e. before the points of coupling $\mu_{\mathrm{c}}$ (gray region) and instability (red region) $\mu^{\star}$. The maximum conversion efficiency is $\xi = 20.2\%$ at $\mu = 0.39$}%
\label{fig_03}%
\end{figure}

The work done at resonance frequency excitation is studied along the friction coefficient in Figure~\ref{fig_03} for horizontal forcing. The friction work remains positive and the damper work remains negative for both modes throughout the complete parameter range. However, the excitation work changes sign for the second mode at $0.261 \leq \mu \leq 0.452$. As a result, energy harvesting is possible in this friction value range. 

\section{Conditions for energy harvesting from flutter-type systems}

\begin{figure}[h!]
\centering
\includegraphics[width=1.0\columnwidth]{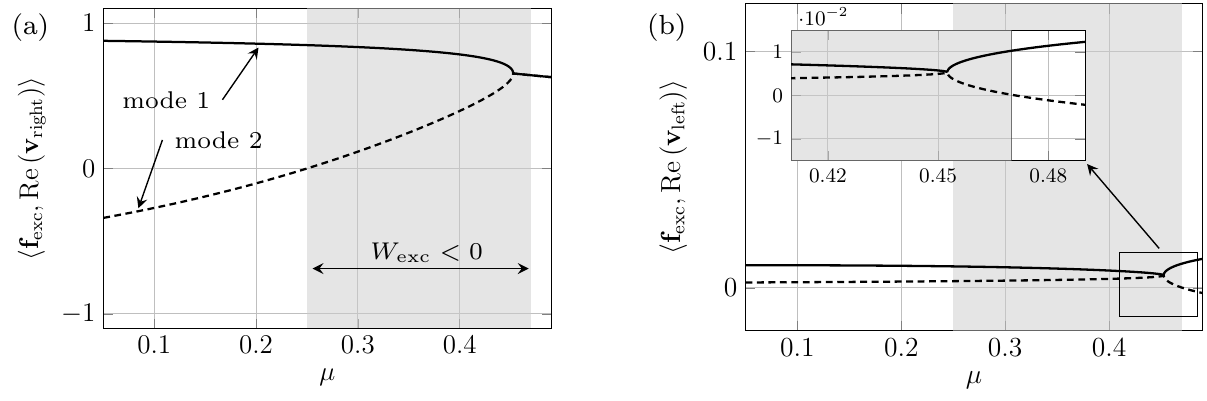}%
\caption{Dot product $\langle\,\cdot\,,\,\cdot\,\rangle$ of the excitation vector $\left[ f_{\mathrm{x}}, f_{\mathrm{y}}\right]^\top$ and real parts of (a) right and (b) left eigenvectors $\mathbf{v_{\mathrm{right, left}}}$. Negative excitation work is observed for the parameter range in between points of orthogonality between forcing and eigenvectors of the second mode}%
\label{fig_04}%
\end{figure}

To better understand the phenomenon of negative excitation work, the relationship between the excitation direction and the eigenvectors of the system is studied in Figure~\ref{fig_04}. The dot product is analyzed to indicate points of orthogonality and co-linearity. For the real parts of the eigenvectors, we observe a relation between the dot product and negative excitation work: There are two parameter configurations at which the forcing acts orthogonal to an eigenvector: at $\mu=0.261$ the forcing is perpendicular to the right eigenvector, whereas at $\mu=0.452$ the forcing is perpendicular to the left eigenvector \cite{Bucher.1997} of the second mode. In between those points, there exists no negative dot product between eigenvectors and forcing. This indicator for energy harvesting was also found in different forcing scenarios. Particularly, negative work is observed for that mode which a) exhibits the change of sign in the dot product and b) later becomes unstable, see Figure~\ref{fig_01}. This behavior needs to be investigated further in future research. Hence, for the time being, we conclude that energy can be harvested from a specific system mode once the dot product of the forcing and all real eigenvectors of this mode share the same sign.     

\section{Conclusion}

We present preliminary results on a non-conservative flutter-type friction oscillator with forcing. In specific configurations, the harmonic forcing allows to pull energy out of the system and thus harvest energy. While the action of the dampers remains unchanged, the selection of the excitation direction changes the qualitative effect of the forcing on the system's energy balance: The flow of energy can be changed from \textit{two sources - one sink}, i.e. positive friction and excitation contributions, to \textit{one source - two sinks}, i.e. positive friction and negative excitation contributions, for energy production.  First results indicate that the forcing direction and the set of left and right eigenvectors have to share a dot product of the same sign for energy harvesting. Particularly, the system is considered in the linearly stable regime, which creates the \textit{energy-on-demand} nature of this approach with two major advantages. First, the energy harvesting is controlled completely by the forcing and therefore does not depend on the existence of ambient or friction-excited vibrations. Once energy is required, the forcing is enabled. Second, also the amplitude and frequency of vibration is controlled by the forcing and can thus be adapted to operational conditions or structural requirements of the system. Once the forcing is turned off, the oscillator returns to its equilibrium. This operation principle may be useful for decentralized sensor energy supply at remote locations within complex machines, such as turbo-machinery, by making use of rotating parts.

\section*{Acknowledgment}
M.S. was supported by the German Research Foundation (DFG) within the Priority Program "`calm, smooth, smart"' under the reference Ho 3851/12-1.

\section*{Author contributions}
MS, MT and NH designed the research, MS performed the research and created the artwork, MS and MT wrote the manuscript and NH revised the manuscript.

\appendix

\section{Equations of motion}
\label{EoM}

\begin{align}
\mathbf{M} = 
\begin{bmatrix}
m & 0 \\
0 & m
\end{bmatrix}, \quad
\mathbf{C} =
\begin{bmatrix}
c_{\mathrm{x}} & 0 \\
0 & c_{\mathrm{y}}
\end{bmatrix}, \quad
\mathbf{N} = 
\begin{bmatrix}
0 & \frac{1}{2}\mu k_{\mathrm{y}} \\
- \frac{1}{2}\mu k_{\mathrm{y}} & 0 
\end{bmatrix},  \\ \notag
\mathbf{K} = 
\begin{bmatrix}
k_{\mathrm{x}} + \frac{1}{2}k & -\frac{1}{2} k + \frac{1}{2} \mu k_{\mathrm{y}} \\
-\frac{1}{2} k + \frac{1}{2} \mu k_{\mathrm{y}}& k_{\mathrm{y}} + \frac{1}{2} k
\end{bmatrix}, \quad
\mathbf{f}_{\mathrm{exc}} = 
 \begin{bmatrix}
f_{\mathrm{x}} \cdot \cos \left( \Omega t +\phi_{\mathrm{x}} \right) \\
f_{\mathrm{y}} \cdot \cos \left( \Omega t +\phi_{\mathrm{y}} \right) 
\end{bmatrix}
\end{align}




\end{document}